\documentclass{jfm}
\usepackage{epsfig}
\usepackage{amsmath}
\usepackage{amssymb}
\usepackage{times}
\usepackage{natbib}
\usepackage{bm}

\title[Ordered undulations and undulation chaos]{Competition and bistability of ordered undulations and undulation chaos in inclined layer convection}

\author[K. E. Daniels, O. Brausch, W. Pesch, and E. Bodenschatz]{Karen E. Daniels$^{1,a}$, Oliver Brausch$^2$, Werner Pesch$^2$,
 Eberhard Bodenschatz$^{1,b}$}

\affiliation{$^1$Laboratory of Atomic and Solid State Physics,
Cornell University, Ithaca, NY 14853;
$^2$Physikalisches Institut der Universit\"at Bayreuth, 95440
Bayreuth, Germany, $^a$Current address: Department of Physics, North Carolina State University, 
Raleigh, NC 27695, $^b$Current address: 
Max Planck Institute for Dynamics and Self-Organization, G\"ottingen, Germany}

\date{\today}

\begin{document}

\maketitle

\begin{abstract}
Experimental and theoretical investigations of undulation patterns in high-pressure, inclined layer gas convection at a Prandtl number near unity are reported.  Particular focus is given to the competition between the spatiotemporal chaotic state of undulation chaos and stationary patterns of ordered undulations. In experiments a competition and bistability between the two states is observed, with ordered undulations most prevalent at higher Rayleigh number. The spectral pattern entropy, spatial correlation lengths, and defect statistics are used to characterize the competing states.  The experiments are complemented by a theoretical analysis of the Oberbeck-Boussinesq equations. The stability region of the ordered undulation as a function of their wavevectors and the Rayleigh number is obtained with Galerkin techniques. In addition, direct numerical simulations are used to investigate the spatiotemporal dynamics. In the simulations both ordered undulations and undulation chaos were observed dependent on initial conditions. Experiment and theory are found to agree well.
\end{abstract}


\section{Introduction}

Many spatially extended pattern-forming systems show non-transient steady states in which spatial and temporal correlations fall off quickly; such states are often described as exhibiting ``spatiotemporal chaos'' (STC). Due to the coupling of spatial and temporal degrees of freedom, STC is richer than the purely temporal chaos observed in many low-dimensional nonlinear systems. While spatiotemporally chaotic phenomena have received a great deal of attention in recent years, our understanding of STC is far from satisfactory \citep{Cross-1994-SC, Gollub-1994-SC, Egolf-1998-ILP, Egolf-2000-MES, Gollub-2000-NDC, Bodenschatz-2000-RDR, Daniels-2002-DTI}. One important issue is to reveal universal mechanisms underlying STC phenomena in quite diverse systems. In this paper, we focus on striped patterns observed, for instance, in the shifting patterns of sand dunes, in cloud street formations, and in many biological systems \citep{Cross-1993-PFE}. A well-investigated manifestation of STC in this setting is commonly described as defect turbulence, in which the creation, annihilation, and motion of topological point defects continually change the local wavenumber and orientation of the stripes \citep{Ramazza-1992-STD, Rehberg-1989-TWD, Bodenschatz-2000-RDR, Daniels-2002-DTI, Young-2003-PHD}.  Another feature is the bistability and competition between STC and well-ordered structures. Both may coexist in the same experimental snapshot, while in other cases the system switches between the two chaotically.  While such a scenario seems to be common in many pattern forming systems, only a few cases have been investigated in detail \citep{Cakmur-1997-BCS, Echebarria-2000-SOH}.

\begin{figure*}
\centerline{\epsfig{file=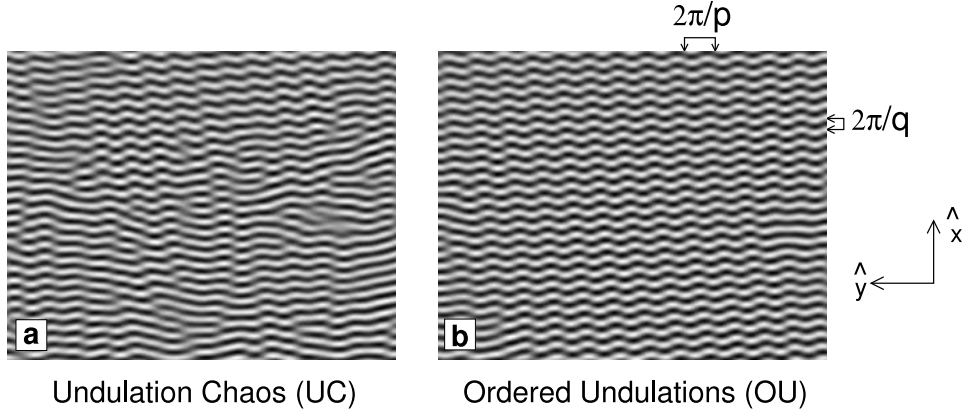, width=5in}}
\caption{Sample shadowgraph images of a convection cell inclined by an angle
$\gamma = 30^\circ$ and at Prandtl number $P = 1.1 \pm 0.04 $ with reduced driving $\epsilon = 0.17$ above onset. Uphill is left; the images have been Fourier-filtered to remove higher harmonics. (a) Undulation chaos (UC) and (b) ordered undulations (OU) \citep{Daniels-2002-DTI,movie}}
\label{f_orderchaos}
\end{figure*}

Here, we study anisotropic stripe (roll) patterns in thermal convection of an inclined fluid layer \citep{Daniels-2000-PFI}. In particular, we present a detailed experimental study of the bistability between {\em ordered undulations} (OU) of stripes and the defect turbulent state of {\em undulation chaos} (UC)) (see Fig.~\ref{f_orderchaos}). The analysis is supported by numerical simulations \citep{Brausch-thesis} of the full Oberbeck-Boussinesq equations (OBE) where the OU and UC attractors were explored in a controlled manner.  We characterize these two states in terms of the spectral pattern entropy, spatial correlation length, and defect density. These measures are found to be correlated with each other and are well suited to distinguish UC from OU. Along with these standard measures of STC, we determine the local wavevector of the patterns and describe the undulations of the stripes in terms of three characteristic amplitudes. These additional quantities allow for a refined discrimination of OU and UC and fit well into the theoretically determined stability island of undulations.

\section{Inclined Layer Convection}
Rayleigh-B\'enard convection (RBC, the thermal instability of a thin horizontal fluid layer heated from below) and its variants have been particularly fruitful in the investigation of pattern formation in extended systems \citep{Bodenschatz-2000-RDR}. Experimentally, the use of compressed gases has allowed for the construction of shallow convection cells with large lateral extent and fast time scales \citep{deBruyn-1996-ASR,Bodenschatz-2000-RDR}. In addition, the underlying fluid dynamical equations (Oberbeck-Boussinesq) are well-established and numerical techniques have been developed to simulate systems with many convection rolls \citep{Pesch-1996-CSC,Bodenschatz-2000-RDR}.

\begin{figure}
\centerline{\epsfig{file=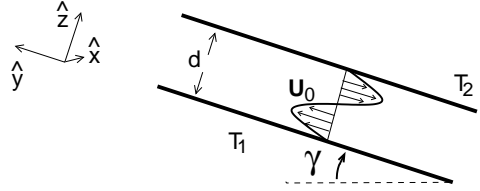, height=1in}}
\caption{Schematic of a convection cell of thickness $d$ inclined by an angle $\gamma$ subject to a temperature difference $\Delta T = T_1-T_2$ with associated coordinate system. $\bm U_0$ indicates the induced shearflow (see text).}
\label{f_schematic}
\end{figure}

One natural variation of RBC is inclined layer convection (ILC), in which the thin fluid layer is additionally tilted by an angle $\gamma$ with respect to horizontal, making the system anisotropic. This situation is common in nature as convective systems are often inclined with respect to gravity. As shown schematically in Fig.~\ref{f_schematic}, the component of gravity parallel to the layer generates a cubic shear flow profile, $\bm U_0(z)$, upwards along the hotter (lower) plate and downward along the cooler (upper) plate. Therefore, ILC may also serve as a prototype system for convective systems in the presence of shear flow.

\begin{figure}
\centerline{\epsfig{file=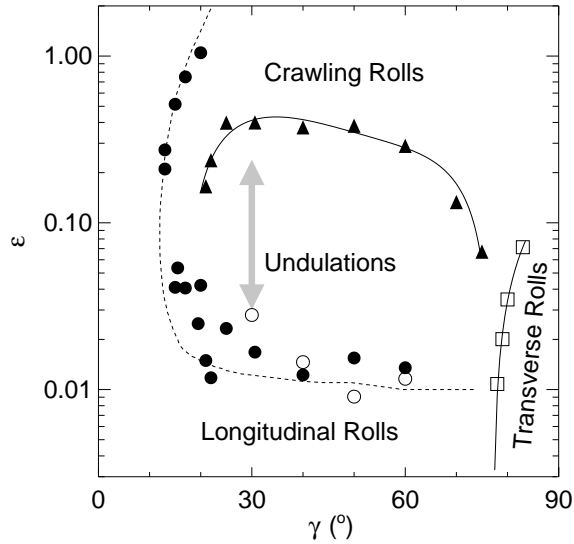, width=3in}}
\caption{Phase diagram for inclined layer convection at $P=1.07$, after \protect\citep{Daniels-2000-PFI}. Data points (and solid lines as guides to the eye) are observed boundaries between the different morphologies. Dashed line is Galerkin prediction for the instability of LR to OU. Gray arrow represents the region of data collection in this paper. }
\label{f_phaseplot}
\end{figure}

As in RBC, the fluid becomes unstable to convection rolls above a critical temperature difference, $\Delta T_c$. We use the reduced control parameter $\epsilon \equiv \frac{\Delta T}{\Delta T_c} - 1$ to measure the distance from the primary instability for fixed angle of inclination $\gamma$. For a Prandtl number $P = \nu/\kappa \approx 1$ (with kinematic viscosity $\nu$ and thermal diffusivity $\kappa$) both buoyancy and shear driven instabilities are observed, which evolve into numerous spatiotemporally chaotic states \citep{Daniels-2000-PFI}. In Fig.~\ref{f_phaseplot}, we have reproduced the part of the phase diagram relevant for the patterns examined in this paper.

At small angles of inclination, buoyancy provides the primary instability and the convection rolls are aligned with the shear flow direction (longitudinal rolls, LR) \citep{Clever-1973-FAL}. Above a critical angle $\gamma_{c2}$ ($\approx 78^\circ$ for the parameter regime presented here), the primary instability is shear flow driven and the rolls align perpendicularly to the shear flow direction (transverse rolls, TR) \citep{Hart-1971-SFD, Hart-1971-TWV, Clever-1977-ILC, Ruth-1980-TTR, Ruth-1980-SII}.

Over a range of intermediate angles ($15^\circ \lesssim \gamma \lesssim 70^\circ$), transverse modes trigger a secondary bifurcation of LR to three dimensional undulation patterns \citep{Clever-1977-ILC} slightly above onset (see the dashed line in Fig.~\ref{f_phaseplot} at $\epsilon \approx 0.015$.) A tertiary instability to a state of crawling rolls (at $\epsilon \approx 0.3$) limits the existence region of undulations from above. For a more detailed discussion of the phase diagram, see  \citet{Daniels-2000-PFI}.

Experimentally, we observed chaotic undulation patterns in the regime between the lower dashed and the upper solid destabilization line (``undulation chaos''), where undulating convection rolls perpetually break and reconnect via moving point defects. Here, we restricted our investigations to a fixed angle of $\gamma = 30^\circ$ for a range of $\epsilon$ well inside the stability regime shown in Fig.~\ref{f_phaseplot}.

ILC can serve as a paradigm for a class of anisotropic pattern-forming systems such as liquid crystal convection \citep{Kramer-1995-CIN}, Taylor-Couette flow \citep{Tagg-1994-CTP}, annular convection \citep{Kurt-2004-HCR}, sand ripples \citep{Blondeaux-1990-SRU, Hansen-2001-SBT}, or optical pattern formation \citep{Ramazza-1992-STD} which also exhibit defect turbulence. In addition, it may relate to shear flow driven instabilities as observed in cloud street formation \citep{Kelly-1994-ODT}, Taylor-Couette flow \citep{Andereck-1986-FRC}, Poisseuille-B\'enard convection \citep{Kelly-1977-OND, Yu-1997-SMT, Muller-1992-TCP}, or turbulent bursting in Couette flow \citep{Bottin-1998-DTS}.


\section{Experiment} \label{sec:exp}
We conducted our experiments in a thin layer of compressed CO$_2$ within a rectangular cell of height $d = (388 \pm 2) \mu$m and dimensions ($100 \times 203)d$. The CO$_2$ gas was 99.99\% pure and the pressure was $(56.5 \pm 0.01)$ bar regulated to $\pm 55$ mbar. The mean temperature of the convection cell was held constant at $(28 \pm 0.05) ^\circ$C regulated to $\pm 0.3$ mK. For these conditions, the Prandtl number was $P = 1.1 \pm 0.04$ as determined from a materials property program \citep{deBruyn-1996-ASR}. The characteristic vertical diffusion time was $\tau_v = d^2/\kappa = (1.532 \pm 0.015)$ sec. The experiments were conducted at a fixed inclination angle of $\gamma = (30.00 \pm 0.02)^\circ$. We obtained images of the convection pattern using a digital CCD camera, via the usual shadowgraph technique \citep{deBruyn-1996-ASR,Trainoff-2002-POT}.

We collected data at 17 equally spaced values of $\epsilon$ between 0.04 and 0.22, each reached by quasistatic temperature increases from below. At each $\epsilon$, we recorded at least 400 series of 100 images (at 3 frames per second), with each series separated from others by at least $100\tau_v$ to guarantee statistical independence. These runs are later referred to as  short runs.
For the lowest value of $\epsilon=0.04$, we recorded up to 600 series to cope with the reduced number of defects present in that regime. In addition, we collected
data while decreasing the temperature quasistatically from $\epsilon = 0.12$ to $\epsilon = 0.06$ to check for hysteresis, which was not observed. Particularly long time series were recorded at $\epsilon = 0.08$ (one run) and 0.17 (two runs). These consisted of six-hour ($ 1.4 \times 10^4 \tau_v$) continuous runs of data at 3 frames per second. These runs are later referred to as long runs.

\begin{figure*}
\centerline{\epsfig{file=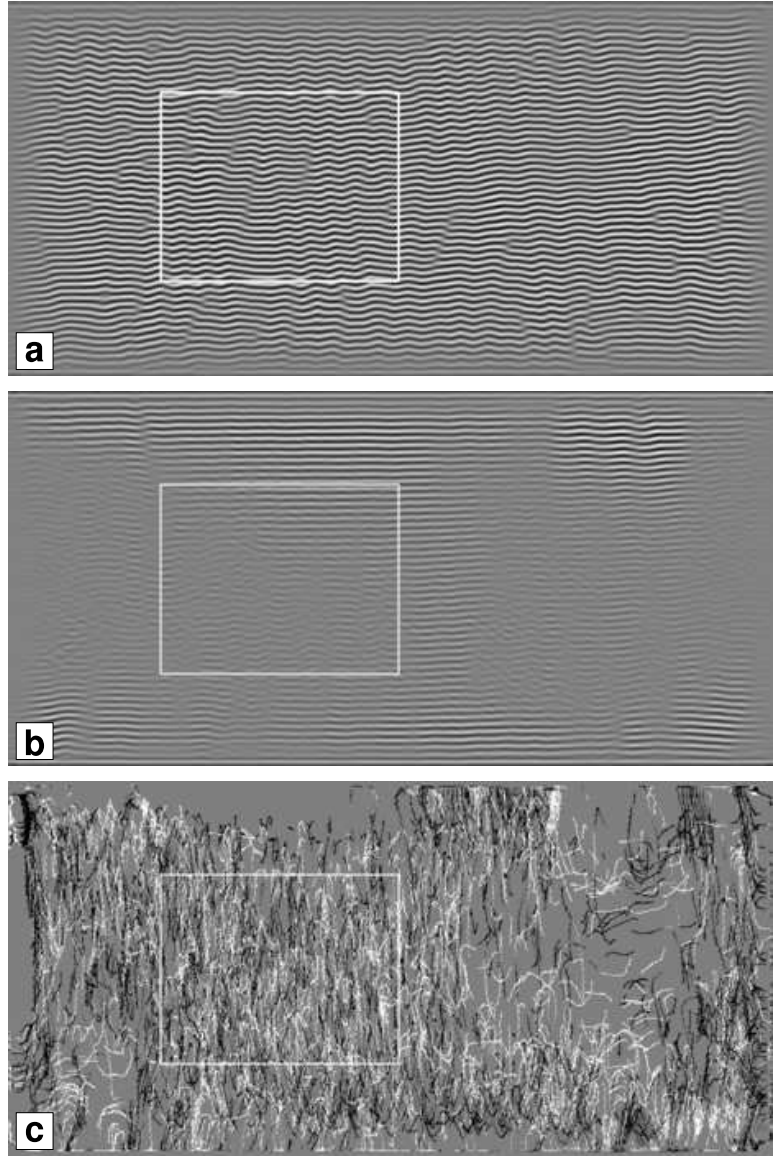, width=4in}}
\caption{(a) Example convection cell, (b) average of 400 statistically independent images, and (c) defect trajectories for fifty statistically independent short runs. White (black)  are defects with positive (negative) topological charge. The nearly homogeneous subregion ($51d \times 63d$) used in all analyses is marked by a white box. Uphill is at the left. $\epsilon = 0.08$, $P = 1.1 \pm 0.04$ and $\gamma = 30^\circ$. }
\label{f_homog}
\end{figure*}


Figure~\ref{f_homog}a shows a sample Fourier-filtered shadowgraph image of the complete experimental cell at $\epsilon = 0.08$. While on first sight the pattern appears to be uniformly composed of UC, detailed investigation demonstrates that this is not the case. In Fig.~\ref{f_homog}b we have superimposed 400 statistically independent images at the same $\epsilon$ (one from each short run). Clearly visible are gray regions, such as within the white-framed subregion, where the average looks almost structureless. There exist also long-term ordered regions such as the strong black/white roll patches near the boundaries. Other spatiotemporally chaotic systems \citep{Ning-1993-STA,Gluckman-1993-TAC} also exhibit similar behavior due to the presence of boundaries. 

Another method of characterizing the spatiotemporal homogeneity of a ``chaotic'' pattern is to superimpose defect trajectories obtained for statistically independent frames (see Section~\ref{sec:defects} for details on defect-detection). An example is shown in Fig.~\ref{f_homog}c, where we superimpose defect trajectories for 50 statistically independent short runs. Defects with positive (negative) topological charge are white (black). Again, far away from the boundaries (within the marked subregion), persistent and homogeneous creation and annihilation of point defects was observed.  In contrast, other regions were either rarely traversed by defects or were preferentially traversed by defects of like charge. In addition, favored spots along the boundaries preferentially seeded defects of the same charge, which then followed similar trajectories. Thus, to minimize the impact these effects and utilize as ideal a cell as possible, the data presented in this paper were taken only from the subregion of dimension $51d \times 63d$ marked by the white box.

\section{Theoretical description }\label{sec:theor}

We base our theoretical description on the standard Oberbeck-Boussinesq equations (OBE), slightly generalized to cover the inclination of the fluid layer. As in the experiments, we concentrate on systems with large aspect ratio, {\em i.e.}, with the lateral  dimensions of the cell in the $x,y$ plane $\gg$ its height $d$ in the $z$-direction. Thus, we adopt the usual idealization of periodic horizontal boundary conditions, which are incorporated numerically by switching from position space to 2-dimensional Fourier space.

First, the onset of ILC is obtained by a standard linear stability analysis of the base state. Then, in the nonlinear regime, we investigate the stability of periodic solutions using both weakly nonlinear and multimode Galerkin approaches. Most important for the interpretation of the (weakly) disordered states in the experiments have been fully three-dimensional solutions of the OBE.

\subsection{Basic equations}

As shown in Fig.~\ref{f_schematic}, the convection cell is subjected to a temperature difference $\Delta T \equiv (T_1 - T_2) > 0$ between the hot (bottom) and cold (top) plate and is inclined by an angle $\gamma$. In a coordinate system aligned with the cell, the gravity vector is given by ${\bf g} = -g \left( {\bf\hat z} \cos\gamma + {\bf\hat y} \sin\gamma \right)$, where $g$ is the acceleration due to gravity.
As usual we introduce the Rayleigh number $R$ as a nondimensionless measure for $\Delta T$:
\begin{equation}
R \equiv \frac{\alpha g \cos \gamma \Delta T d^3}{\nu \kappa}
\label{e_Rdef}
\end{equation}
with $\alpha$ the thermal expansion coefficient, $\nu$ the kinematic viscosity, $\kappa$ the thermal diffusivity. Please note that $R$ depends on $\gamma$.

The OBE become dimensionless if lengths are measured in units of $d$, time in units of the thermal diffusion time $\tau_v \equiv d^2/\kappa$, temperatures in units of $\Delta T/R$ and velocities in units of $d/\tau_v$.
Thus, we arrive at the following nondimensional form of the OBE:
\begin{subequations}\label{eq:obe}
\begin{align}
\nabla^2{\bf u}+ \theta \, {\bf\hat z} + (-Rz + \theta) \, {\bf\hat y}\tan{\gamma}
-{\bf \nabla} \pi \;
& = 
\; P ^{-1} \left[ \frac{\partial}{\partial t}{\bf u}+({\bf u
\cdot \nabla}) {\bf u} \right],
\label{e_OBE1} \\
\nabla^2 \theta + R{\bf u \cdot \hat z} \;
& = 
 \;\frac{\partial\theta}{\partial t} + ({\bf u \nabla} )\theta
\label{e_OBE2}.
\end{align}
\end{subequations}

$ (\Delta T/ R) \theta$ denotes the deviation from the linear temperature profile $\frac{1}{2}(T_1 + T_2) - \Delta T \, z$ ($-1/2 < z < 1/2$) of the basic non-convecting state. Terms which can be expressed as gradients are included in the pressure term $\nabla \pi$. We assume incompressibility of the velocity field, $\bf u$, i.e. ${\bf \nabla \cdot u} = 0$. Furthermore $\theta$ and $\bf u$ are required to vanish at
$z = \pm 1/2$ (rigid boundary conditions). 

Even below the onset of convection, the OBE provide a base flow with a cubic mean-flow profile ${\bf U}_0(z)$ (in dimensionless units):
\begin{equation}
{\bf {U}}_0(z) = \frac{R \tan 
\gamma}{6}
\left( z^3-\frac{z}{4} \right) \hat{\bf y},
\label{e_meanflow}
\end{equation}
as illustrated in Fig.~\ref{f_schematic}. Above the onset of convection, it is useful to split the solenoidal velocity field $\bm u$ into three parts using 
\begin{equation}\label{eq:solen}
{\bf u} \equiv {\bf U}_0(z) + {\bf v}(x,y,z;t) + {\bf U}(z,t).
\end{equation}
The term ${\bf v}(x,y,z;t)$ is derived from a poloidal-toroidal decomposition of $\bf u$ in the form ${\bf v} = {\bf v}_{pol} + {\bf v}_{tor} \equiv \nabla \times \nabla \times({\bf \hat z} \chi ) + \nabla \times \zeta$ with the poloidal (and toroidal) velocity potentials $\chi$ (and $\zeta$). This decomposition  automatically fulfills  ${\bf \nabla \cdot v} = 0$. The field ${\bf v}_{pol}(x,y,z;t)$ describes the periodic convection rolls with their (nearly circular) streamlines and vanishing vertical vorticity $\Omega_z = (\nabla \times {\bf u})_z$, while ${\bf v}_{tor}$ is associated with finite $\Omega_z$.

Convection also leads to a modification  ${\bf U}(z,t)$ (see Eq.\ref{eq:solen}) of the basic shear flow profile ${\bf U}_0(z)$. The governing equations for ${\bf U}(z,t) = ({U}_x(z,t), U_y(z,t),0)$, derive from a horizontal average of Eq.~\ref{e_OBE1} and read as follows:
\begin{subequations} \label{eq:mean}
\begin{align}
\left( \partial^2_{zz}-P^{-1}\partial_t \right ) {U_x}
+ P^{-1} 
\partial_z \overline{[v_x v_z]}
\;=&\;0,
\\
\left( \partial^2_{zz}-P^{-1}\partial_t \right) {U_y} +
P^{-1}\partial_z \overline{[v_y v_z]} +\overline{\theta}\sin{\gamma}
\;=& \;0,
\end{align}
\end{subequations}
Here, the overline denotes an in-plane average (the ${\bf q} = 0$ component in Fourier-space) of the corresponding terms. ${{\bf U}}(z,t)$ shares the odd $z$-symmetry with  ${\bf U}_0 (z)$, but with the flow direction reversed. Thus, the baseflow ${\bf U}_0 (z)$, which is potentially unstable due to the inflection point at $z =0$, is reduced with increasing $R$. Note that Eqs.~\ref{eq:obe} become $\gamma$-independent for LR which vary only in the $x-$ direction. Thus the onset of LR and their structure (but not their stability) can be deduced immediately  from the standard RBC problem without inclination \citep{Clever-1973-FAL}.

While ${\bf v}_{tor}$ vanishes in periodic roll patterns, it is excited by any roll curvature or defects in the pattern. In contrast to ${\bf v}_{pol}$, the Hagen-Poisseuille-like $z-$profile of ${\bf v}_{tor}$ is even in $z$, and is often referred to as mean-flow. Other fields can thus be efficiently advected by ${\bf v}_{tor}$, typically leading to further enhancement of some initial perturbation. For Prandtl number of order unity and below, the resulting positive feedback is the source of spatiotemporal complexity as observed in defect turbulence \citep{Daniels-2002-DTI} and spiral defect chaos \citep{Morris-1993-SDC}.

\subsection{Periodic solutions}

It is convenient to substitute Eq.~\ref{eq:solen} into Eqs.~\ref{eq:obe} and to use a condensed notation for the resulting OBE equations:
\begin{equation}
{\mathcal C} \frac
{\partial}{\partial t} {\bf V}(\bm x, z, t) = {\mathcal L} {\bf V} (\bm x, z, t) + {\mathcal N}({\bf V}, {\bf
V}).
\label{e_symbol} \end{equation}
The fields $\chi$, $\zeta$, $\theta$, $\overline{\bm u}$
have been collected into a symbolical vector $\bf V$.
The letters $\mathcal C$ and $\mathcal L$ represent the resulting linear operators and $\mathcal N$ the quadratic nonlinearities in Eqs.~\ref{eq:obe} and \ref{eq:mean}. 

The first step in any theoretical investigation is the {\it linear} stability analysis of the base state ${\bf V} = 0$. Neglecting $\cal N$ in Eq.~\ref{e_symbol}, we arrive at a linear eigenvalue problem which diagonalizes in Fourier space 
via the ansatz ${\bf V}({\bf x},z,t) = e^{\lambda t} e^{i\bf q \cdot x} {\bf V}_{lin}({\bf q},z)$ with 
${\bf x}=(x,y)$ and ${\bf q} = (q, p)$. The eigenvalue $\lambda(\bm q, R)  = \sigma + i \omega $ with the largest real part provides the growth rate $\sigma(\bm q,R)$. The condition $\sigma(\bm q, R) = 0$ describes the neutral stability curve $R_0(\bm q)$ which has a minimum at $R_c = R_0({\bf q}_c)$ for the critical wave vector ${\bf q}_c$. In the present case, we find $\omega=0$, so the bifurcation is stationary.

In contrast to isotropic systems, where $\sigma({\bf q})$ and $R_0({\bf q})$ depend only on the modulus of $\bf q$, the anisotropy causes them to additionally depend on the angle between $\bf q$ and ${\bf \hat y}$. Inspection of Eq.~\ref{eq:obe} shows that the control parameter $R$ appears not only explicitly in the buoyancy term of Eq.~\ref{e_OBE2}, but also in the coefficient $P^{-1} R \tan \gamma$ arising from the ${\bf U}_0$  contributions  (see \ref{e_meanflow}) to  the advection terms (see  Eq.~\ref{eq:solen}).

Where buoyancy prevails ($\gamma$ less than the critical angle $\gamma_{c2}(P)$), we find LR at the onset of convection, with ${\bf q}_c = (q_c, 0)$. The threshold $R_c = 1708$ (defined according to Eq.~\ref{e_Rdef}) and ${\bf q}_c = 3.117$ are $\gamma$-independent. In contrast, for $\gamma > \gamma_{c2}$ the instability is shear-flow driven and TR with ${\bf q}_c = (0,p_c)$ are selected. This critical behavior at threshold was discussed in \citet{Hart-1971-SFD,Hart-1971-TWV,Clever-1973-FAL,Ruth-1980-SII,Clever-1977-ILC}.

We use Galerkin methods to study ideal periodic convection patterns in the {\it nonlinear} regime. All fields are expanded in two-dimensional Fourier series with respect to the $(\bf{ \hat x}, \bf{\hat  y})$-coordinate plane and in suitable test functions in $z$ fulfilling the rigid boundary conditions at $z = \pm 1/2$. For instance, the $\theta$ component of $\bf V$ reads
\begin{equation}
\theta({\bf {x}},z;t)=\sum_{k,m}\,\left[ \exp{[i(k q x + m p y
)]}\,\sum_{n}\, c_{k,m;n}(t)\,S_n(z) \right]
\label{e_ansatztheta} \end{equation}
with $S_n(z) = \sin (n \pi (z +1/2))$. To ensure real-valued $\theta$, we require $c_{-k,-m;n}=c^*_{k,m;n}$. The fields $\chi, \zeta$ are analogously represented except that we use the Chandrasekhar functions $C_n(z)$ \citep{Chandrasekhar-1961-HHS} instead of the $S_n(z)$ for $\chi$.

For LR with ${\bf q} \parallel {\bf\hat x}$ all expansion coefficients vanish for $m \ne 0$ and  conversely for $k=0$ for TR with ${\bf q} \parallel {\bf \hat y}$. By inserting Eq.~\ref{e_ansatztheta} and the analogous expansions for $\chi, \zeta, {\overline {\bf u}}$ into Eq.~\ref{e_symbol} and truncating the series one arrives at a nonlinear algebraic system for the their expansion coefficients. This system of equations is solved using a Newton-Raphson iteration scheme and subsequently tested for stability with respect to linear perturbations. Using the standard Floquet ansatz, the $\theta$ perturbation reads for instance
\begin{equation}
\label{eq:gal}
\delta \theta({\bf x},z;t)= e^{\sigma_n \, t} e^{i \bm s \bm x}
\sum_{k,m} \left[ \exp {[i(k q x + m py )]} \sum_{n} \delta
c_{k,m;n} S_n(z) \right].
\end{equation}
with $\bm s = (s_x, s_y)$.
Positive nonlinear growth rates $\Re e[\sigma_n({\bf q}, \bm s, R, P)]$ signal secondary instabilities of the periodic convection patterns. Numerical results agree with all those published previously in the literature ({\em e.g.} \citet{Clever-1977-ILC}). 


This paper investigates the secondary, modulational, wavy (zig-zag, undulation) instability of LR at $R = R_u(\gamma, P)$. This instability is initially characterized by transverse undulations along the roll axis at large wavelength ($s_x=0$ and $s_y \rightarrow 0$)  as first described in \citep{Ruth-1980-SII,Clever-1977-ILC}. It develops for $R \geq R_u$ slightly above $R_c$, {\em i.e.} at a small reduced control parameter  $\epsilon_u(P, \gamma) \equiv (R_u-R_c)/R_c$. For example,  $\epsilon_u = 0.016$ for $\gamma = 30^\circ$ on the  dashed line in Fig.~\ref{f_phaseplot}.

Weakly nonlinear analysis provides some insight into the underlying physical mechanisms of the instability, particularly the final-amplitude state. We utilize a set of coupled, complex  amplitude equations, which can be expected to work reliably for small $\epsilon$.

We use the following ansatz for the solution of Eq.~\ref{e_symbol}: 
\begin{equation}
{\bf V ({\bm x}, z, t}) = A(t) {\bf V}_{lin}({\bf q}_1, z) e^{i {\bf q}_1 {\bf x}}+ B(t) {\bf V}_{lin}({\bf q}_2, z)e^{i {\bf q}_2 {\bf x}}- C(t) {\bf V}_{lin}({\bf q}_3, z)e^{i {\bf q}_3 {\bf x}}+ \mbox {c.c. + h.o.t.}
\label{e_Vampl}
\end{equation}
with ${\bf q}_1=(q,0)$, ${\bf q}_2=(q,p)$ and ${\bf q}_3=(q,-p)$. The first term describes LR of wavenumber $q$ and complex amplitude $A(t)$, while the additional terms provide transverse modulations with wavenumbers $\pm p$ and complex amplitudes $B(t)$ and $C(t)$. Note that the wavevectors ${\bf q}_i$ span a resonant tetrad obeying
\begin{equation}
{\bf q}_1={\bf q}_2+{\bf q}_3 -{\bf q}_1.
\label{e_wavyres}
\end{equation}
We insert Eq.~\ref{e_Vampl} into Eq.~\ref{e_symbol} and retain terms up to cubic order in the amplitudes $A, B, C$. After projecting these equations onto the linear eigenvectors ${\bf V}_{lin}({\bf q}_i, z)$, we obtain the following coupled amplitude equations, valid for $R \ge R_u$.

\begin{subequations}\label{eq:ampli}
\begin{align}
\frac{d}{dt}A&=\left( \sigma_A
-b_{11}|A|^2
-b_{12}|B|^2
-b_{12}|C|^2
\right) A
+\rho_1A^*BC,
\label{e_amplA}
\\
\frac{d}{dt}B&=\left( \sigma_B
-b_{21}|A|^2
-b_{22}|B|^2
-b_{23}|C|^2
\right) B
+\rho_2AAC^*,
\label{e_amplB}
\\
\frac{d}{dt}C&=\left( \sigma_C
-b_{31}|A|^2
-b_{32}|B|^2
-b_{33}|C|^2
\right) C
+\rho_3AAB^*.
\label{e_amplC}
\end{align}
\end{subequations}

The linear growth rates are defined as: $\sigma_A(R) = \sigma ({\bf q}_1,R), \sigma_B(R) = \sigma_C(R) = \sigma({\bf q}_2,R) < \sigma_A$.  The nonlinear coupling coefficients $b_{ij}>0$ and the cross-coupling coefficients $\rho_i >0$ which characterize a resonant tetrad coupling were calculated using the usual projection techniques.
It is convenient to characterize the complex amplitudes by their moduli and phases:
\begin{equation}
A=|A|\exp{(i \phi_A)}, \quad
B=|B|\exp{(i \phi_B)}, \quad
C=|C|\exp{(i \phi_C)}.
\end{equation}
Two phases can be arbitrarily chosen due to translational invariance in the $x, y$ directions and the choice $\phi_A =0, \, \phi_B =0$ leading to real $A,B$ is convenient. The undulation growth rate is obtained by a linear stability analysis with respect to $B$ and $C$. Solutions with $|A| > 0$ and $|B| = |C| > 0$ exist only if the condition $\sigma_u(R) \equiv \sigma_B(R) + (\sigma_A(R) + \cos(\phi_C) \rho_2 +b_{21})/b_{11} > 0$ holds. $\sigma_u(R)$ is minimal for $\phi_C= 0$  since $\rho_2 >0$ and crosses zero at a value $R = R_u$, which agrees  with the full Galerkin result.

In more physical terms, the undulation solution creates horizontal variations of the mid-plane temperature field, which correspond to the experimentally-obtained shadowgraph pictures. It is convenient to normalize the temperature component of ${\bf V}_{lin}({\bf q}, z)$ to unity at the midplane ($z = 0$). Thus, we arrive from Eq.~\ref{e_Vampl} at the the following representation for the mid-plane temperature $\psi_R$ field of an ideal undulated pattern:
\begin{equation}
\psi_R(x,y) =
\Re \left[ e^{iqx}(A(q,p) + B(q,p) e^{ipy} - C(q,p) e^{-ipy}) + h.o.t.) \right].
\label{e_perfect}
\end{equation}
with real $A,B$ and $C = B$ and higher order terms neglected.
It is illuminating that, slightly above the instability, $\psi_R$ as presented in Eq.~\ref{e_perfect} can be considered  as a phase modulation of the LR  pattern. In fact, a modulated LR pattern $ 2 A \cos(qx +\phi(y))$ with $\phi(y)= \frac{2B}{A} \sin(py)$ and $\frac{2B}{A} \ll 1$ corresponds to Eq.~\ref{e_perfect} for the resonant case $B=C$ in a leading order expansion with respect to $B/A$.

The amplitudes $A$ and $B = C$ obtained from Eqs.~\ref{eq:ampli} and the resulting approximate solution $\bf V$ according to Eq.~\ref{e_Vampl} are convenient for generating the full nonlinear Newton iteration scheme. In this way, we construct fully three-dimensional ordered undulation (OU) solutions of the OBE of the type described in Eq.~\ref{e_ansatztheta} for $R > R_u$. The subsequent linear stability analysis (see Eq.~\ref{eq:gal}) yields the stability regions of the OU.

\begin{figure*}
\centerline{\epsfig{file=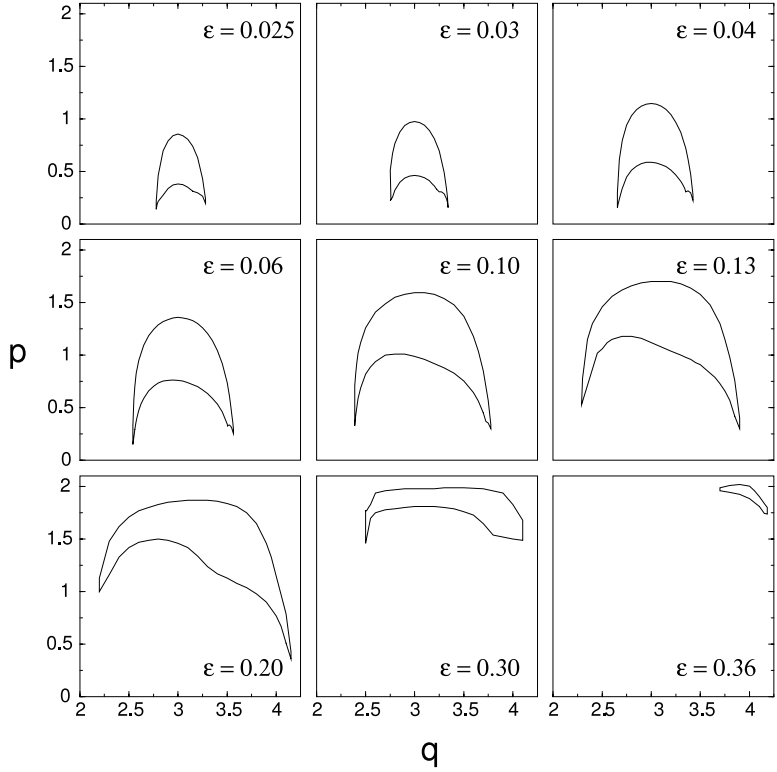, width=4in}}
\caption{Stability islands for ordered undulations at $P=1.14$ and $\gamma=30^\circ$ in the $q,p$ plane at increasing values of $\epsilon$.}
\label{f_moons}
\end{figure*}

Figure~\ref{f_moons} shows the representative results for $P = 1.14$ and $\theta = 30^\circ$. The OU at a given $\epsilon$ are stable within certain regions  (islands) in the $(q,p)$-space. These islands shrink with decreasing $\epsilon$, becoming arbitrarily small as $\epsilon$ approaches $\epsilon_u$. In this limit, the island is located at $q \approx q_c$ and $p \approx 0$ and its area approaches zero. The islands  reach their maximal size for $0.1 \lesssim \epsilon \lesssim 0.2$, before they shrink again as the transition to crawling rolls is approached.

The limits of the islands in Fig.~\ref{f_moons} are given by amplitude instabilities ($\bm s = 0$ in Eq.~\ref{eq:gal}) at the low-$p$ boundary of the islands and long-wavelength modulational instabilities ($ |\bm s | \ll q_c$)  at the high-$p$ boundary. In the former case, the amplitude $A$ is virtually unaffected, whereas the other amplitudes in Eq.~\ref{e_perfect} are perturbed as follows: $B \rightarrow B +\delta$, $C \rightarrow C -\delta$, with $|\delta| \ll B$. Due to the asymmetry $B \ne C$, a non-vanishing component $U_x $ develops, {\itshape i.e.} a mean-flow which likely drives the defect-turbulent state. By analogy to the aligning effect of the basic profile ${\bf U}_0$ on the LR,  $\bm U_x $ will tend to turn the roll axis as well.

In contrast to the amplitude instabilities, the modulational instabilities (which involve finite Floquet vectors $\bf s$) will change the basic periodicity of the undulation
solutions. The main consequence of such a perturbation involves a mean-flow (along ${\bf \hat x}$) contribution to ${\bf v}_{tor}$, which presumably induces defect chaos as well.

\subsection{Simulations}

\begin{figure}
\centerline{\epsfig{file=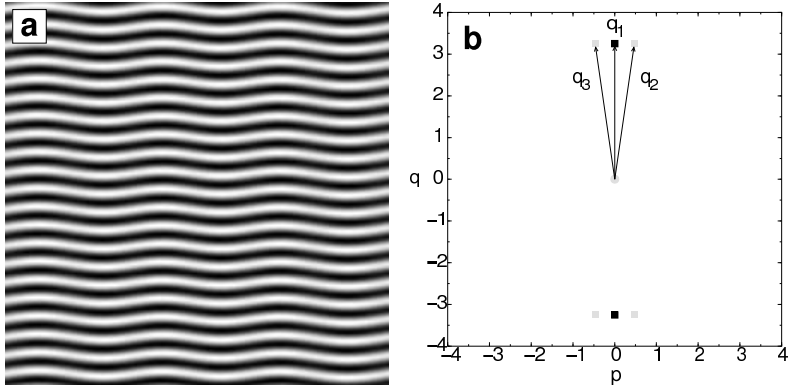, height=2in}}
\caption{Simulation at $\gamma=30$ and $\epsilon=0.025$ ($R=1750$). (a) Stationary OU after $10^5$ iterations ($2000 t_v$), starting from random initial conditions. (b) Fourier spectrum of (a): pattern locked perfectly into the resonant tetrad described by  the amplitude equations.}
\label{fig:wav}
\end{figure}

Since the Galerkin method is restricted to the analysis of stationary periodic states and their stability, direct numerical simulations of the OBE is needed to investigate the time evolution of complex patterns. For that purpose, a  previously-developed  code \citep{Decker-1994-SDC, Pesch-1996-CSC}  was generalized to cover inclination \citep{Brausch-thesis}. This code represents all fields by an appropriate 
Galerkin-ansatz like Eq.~\ref{e_ansatztheta}  and  treats the  $(x,y)$-dependence in Fourier space by a pseudo-spectral technique. By using a small ($< 10$) number of modes in the $\bf\hat z$ direction, reliable simulations of large aspect ratio systems can be conducted on common computer clusters in a reasonable  time. In the $(x,y)$-plane, our simulations covered areas up to $40d \times 75d$, where we used up to $384$ Fourier modes in either direction. Periodic boundary conditions have been used except in a few cases, where we introduced subcritical  $\epsilon$ ramps to  suppress convection at the sidewalls to approximate experimental conditions.  We typically found the same bulk behavior in both cases for simulation domains of size $40d \times 40d$ or greater. 

\begin{figure}
\centerline{\epsfig{file=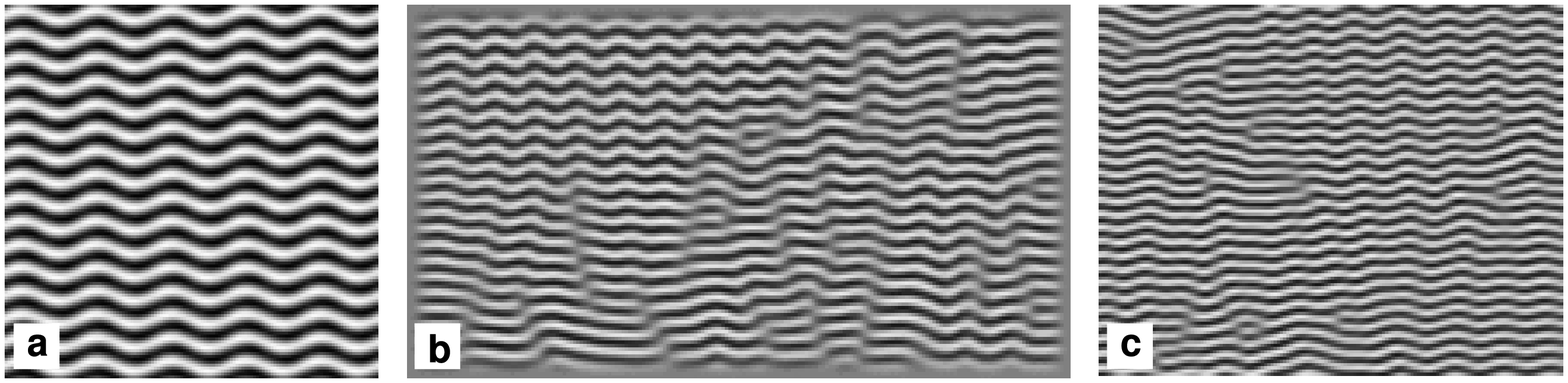, width=\linewidth}}
\caption{Convection patterns at $\epsilon = 0.10$ and $\gamma=30^\circ$. (a) Numerical simulation started from Galerkin solution with superimposed Gaussian noise, showing perfect OU after reaching steady state. (b) Numerical simulation started from Gaussian noise, with subcritical $\epsilon$-ramp simulating lateral boundaries. OU are visible in the upper left and UC in the lower right. (c) Fourier-filtered shadowgraph image of homogeneous subregion of experimental cell.}
\label{f_simexppics} 
\end{figure}

The time integration in the simulations was started  either from random initial conditions (to scan the manifold of nonlinearly selected solutions) or from previously calculated Galerkin solutions with superimposed  noise to check for stability. A representative example of stationary OU is shown in Fig.~\ref{fig:wav}; the system chose wavenumbers $q = 1.05 q_c = 3.26$ and $p= 0.16 q_c = 0.5$, which is inside the corresponding stability island in Fig.~\ref{f_moons}. The resulting ratio $B/A= 0.59$ according to Eq.~\ref{e_perfect} agrees with Galerkin results. In fact, OU were always selected for small $\epsilon < 0.03$ when starting from random initial conditions. At larger $\epsilon$ OU could still be obtained by starting from a stable periodic Galerkin OU  with a small amount of superimposed noise with a signal-to-noise ratio of $10^{-4}$. This is demonstrated in Fig.~\ref{f_simexppics}a. Here we started at $\epsilon = 0.1$  and  $q = 1.05 q_c = 3.28, p = 0.31 q_c = 1.02$ inside the corresponding stability region in Fig.\ref{f_moons} and recovered the ideal Galerkin solution with $B/A = 0.94$. However, the basin of attraction of the OU shrinks with increasing $\epsilon$ and the simulations starting from random initial conditions typically settle into UC, as for example shown in Fig.~\ref{f_simexppics}b. The ensuing bistability between stationary perfectly ordered solutions and dynamic weakly defect turbulent at larger $\epsilon$ is the main theme of this paper.

\section{Analysis of complex undulation patterns}

\begin{figure*}
\centerline{\epsfig{file=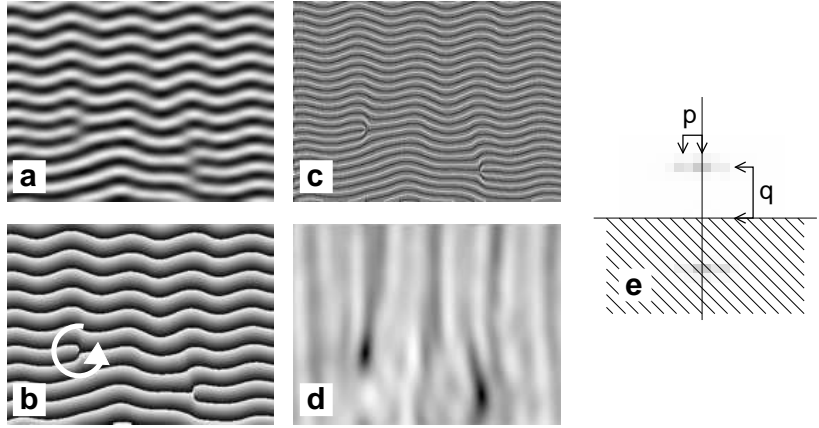, width=0.8\linewidth}}
\caption{Five views of a demodulated experimental convection pattern showing undulations and defects. 
(a) Real part $\psi_R$ of a Fourier-filtered shadowgraph image at $\epsilon = 0.10$ with negative (left) and positive (right) defects. Uphill direction is at the left side of the page. 
(b) Phase field $\phi$: black is $\phi = 0$ and white is $\phi = 2\pi$. The arrow indicates  an integration loop around a defect as given in Eq.~\ref{e_defint}.
(c) Zero crossings of $\psi_R$ and $\psi_I$: white is $\psi_R = 0$ and black is $\psi_I = 0$. 
(d) Modulus of image, $|\psi|$. 
(e) Power spectrum of $\psi_R$. $(q,p)$ are wavenumbers associated with $({\bf \hat x}, {\bf \hat y})$ directions. Shaded region of Fourier space ($q < 0$) is discarded during reconstruction of the complex field.}
\label{f_pics}
\end{figure*}

In this section we discuss several approaches  to  characterizing STC in general and the onset of chaotic undulations in particular. Section~\ref{sec:defects} explains how we identify all topological defects in a shadowgraph image, and generate a time series of the total number of defects, $N(t)$, from the consecutive images of an experimental run. Section \ref{sec:stoch} deals with the pattern entropy  $S(t)$ and the transverse correlation length $\xi (t)$. Finally, in Sec. \ref{sec:locamp} we discuss the statistical properties of the amplitudes $A, B, C$, defined in Eq.~\ref{e_Vampl}.

\subsection{Identification of defects \label{sec:defects}}

A shadowgraph (or simulation) image can be understood as the real part $\psi_R$ of a complex field $\psi(x,y) = |\psi(x,y)| \exp[i\phi(x,y)]$ with modulus $|\psi|$ and phase $ \phi$. To construct $\psi(x,y)$, we demodulate the image as illustrated in Fig.~\ref{f_pics}: after a 2d Fourier transformation, half of the modes in the Fourier plane are set to zero (see Fig. \ref{f_pics}e). A subsequent inverse Fourier transformation recovers $\psi(x,y)$.

As already discussed above, the characteristic feature of a UC pattern is undulating stripes containing topological point defects. Defects locally change the spacing of rolls and their orientation \citep{Bodenschatz-1988-SDD}. Complex analysis requires that defects be located at a zero of $\psi(x,y)$: {\em i.e.} at the points where the lines of zero real part, $\psi_R(x,y) = 0$, and zero imaginary part, $\psi_I(x,y) = 0$, cross \citep{Rasenat-1990-ESD}. (See Fig.~\ref{f_pics}bc for an illustration.) As a second criterion, the topological charge (or ``winding number''), $n$, defined by the circulation of the phase gradient about a defect 
\begin{equation}
\oint \vec{\nabla} \phi \cdot d \vec{s} = \pm n 2 \pi
\label{e_defint}
\end{equation}
becomes nonzero; only point defects with $|n|=1$ are found in UC. 

To locate the zero crossings efficiently \citep{Egolf}, we formed a set of four binary fields for $\psi_R(i,j)$  at each pixel ($(i,j)\in \mathbb{Z}$) of the plane.
\begin{equation}
\tilde{\psi}_R^{\Delta i, \Delta j}(i,j) =
\begin{cases}
1, & \psi_R(i+\Delta i, j + \Delta j) > 0 \\
0, & \text{otherwise}
\end{cases}
\end{equation}
where $(\Delta i, \Delta j) \in \{(0,0), (0,1), (1,0), (1,1)\}$ denote the neighboring pixels of $(i,j)$. The four binary fields $\tilde{\psi}_R^{(i,j)}$ are then logically combined at each point to produce a new binary field $\overline{\psi}_R(i,j)$.
\begin{equation}
\overline{\psi}_R = (\tilde{\psi}_R^{0,0} \oplus \tilde{\psi}_R^{0,1}) \;
\vee \; (\tilde{\psi}_R^{0,0} \oplus \tilde{\psi}_R^{1,0}) \;
\vee \; (\tilde{\psi}_R^{0,0} \oplus \tilde{\psi}_R^{1,1})
\end{equation}
where $\oplus$ is the XOR (exclusive OR) logical operator and $\vee$ is the OR logical operator. Analogously, we define the new binary field $\overline{\psi}_I(i,j)$ as well. All points $(i,j)$ where $\overline{\psi}_R(i,j) = \overline{\psi}_I(i,j) = 1$ serve as candidates for a defect. To ensure that such a point is indeed a defect, the contour integral in Eq.~\ref{e_defint} is evaluated by traversing its eight nearest neighbors. A defect candidate is accepted if the contour integral yields $\pm 2 \pi$, assigning the appropriate sign to the defect based on the sign of the integral. After locating all defects, we eliminated double-detections. Counting these zeros of $\psi(x,y)$ for a time series of frames yields $N(t)$.

\subsection{Global stochastic properties of undulation chaos}
\label{sec:stoch}
The following section continues to analyze the representative case of inclination angle $\gamma = 30^\circ$.  A prominent feature of the experiments is that at higher $\epsilon \gtrsim 0.10$ the competing attractors -- OU with few defects and UC with many as shown in Fig.~\ref{f_orderchaos}  -- are alternately visited in time, while the system remains persistently chaotic at lower $\epsilon$. 
A similar bistability was also found in the case of (isotropic) RBC: \citet{Cakmur-1997-BCS} observed competition between ideal rolls (IR) and the spatiotemporally chaotic state of spiral defect chaos (SDC). In that case, however, a transition from SDC to IR was observed by decreasing $\epsilon$. There, the analysis of the spectral entropy $S(t)$ \citep{Neufeld-1994-PFR} was illuminating, and the concept is applied here as well. $S(t)$ is defined as 
\begin{equation}
S(t) = - \langle P(q,p,t) \ln P(q,p,t) \rangle
\label{e_entropy}
\end{equation}
where $P(q,p,t)$ is the normalized spectral distribution function that describes the power in the mode with wavevector $(q,p)$ at time $t$ and the average $\langle \cdot \rangle$ is performed over the $(q,p)$ plane.

The function $S(t)$ provides a measure for order in a pattern: we have $S = \ln 6$ for OU described by the six modes of Eq.~\ref{e_perfect}  while a disordered pattern has more modes excited and $S > \ln 6$. Conceptionally for anisotropic patterns, like undulations, the pattern entropy is even better suited to characterize  disordered  pattern than in isotropic systems. For instance a target pattern 
is represented by a ring Fourier space and therefore has a deceptive high value $S$,  
though it is a well ordered state. 

\begin{figure*}
\centerline{\epsfig{file=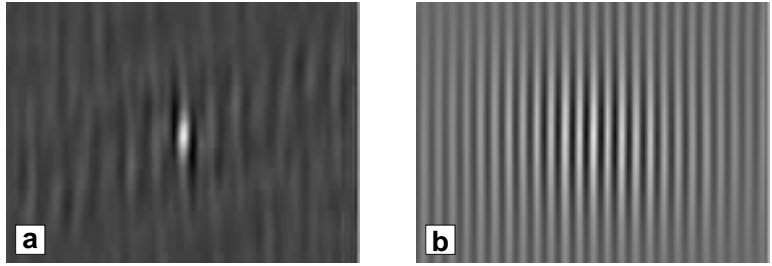, width=4in}}
\caption{ Spatial autocorrelations $\langle M({\bf r}) M({\bf r} -{\bf r}_0) \rangle$ of the modulus $M = |\psi|$ of the complex demodulation functions $\psi$ associated with the images shown in Fig.~\ref{f_orderchaos}. (a) UC and (b) OU.  Images are rotated by 6$^\circ$ clockwise with respect to the coordinate system shown in Fig.~\ref{f_orderchaos}, due to a slight misalignment of the experimental cell.}
\label{f_spatcorr}
\end{figure*}

Further characterization of the relative order of the pattern is possible using the correlation length $\xi$ in the transverse ($\hat \bf y$) direction obtained from the half-width at half-maximum of the spatial autocorrelation functions, as demonstrated  in Fig.~\ref{f_spatcorr}. In image (a), the pattern is disordered, the spatial autocorrelation function is  sharply peaked at the origin, and $\xi$ only a few $d$. In image (b), the pattern is highly ordered and shows correlations that fall off with a correlation length approaching the length of the cell.

\begin{figure*}
\centerline{\epsfig{file=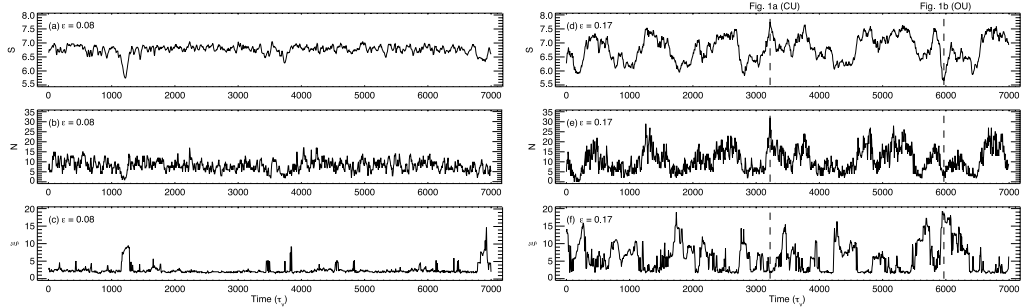, width=\linewidth}}
\caption{Time trace of (a,d) spectral entropy $S$, (b,e) number of defects $N$, and (c,f) transverse correlation length $\xi$ at $\epsilon = 0.08$ (a,b,c) and $\epsilon = 0.17$ (d,e,f), extracted from shadowgraph images. Data was sampled every 10 seconds ($6.5 \tau_v$).}
\label{f_order}
\end{figure*}

Fig.~\ref{f_order} shows example time traces of $N(t)$, $S(t)$ and $\xi(t)$ for $\epsilon = 0.08$ in the UC regime and also for $\epsilon = 0.17$, well above the transition to competition between UC and OU. The dashed vertical lines for the case $\epsilon = 0.17$ mark the times of the images shown in Fig.~\ref{f_orderchaos}, which exemplify the cases of high $S$ (disordered UC) and low  $S$ (ordered OU), respectively.  For each time series  of the quantities $X = S(t), N(t), \xi(t)$ we obtain the time average $\langle  X \rangle $ and the standard deviations $\Sigma_X \equiv \sqrt{\langle {X^2} \rangle - \langle X \rangle^2}$.

\begin{figure*}
\centerline{\epsfig{file=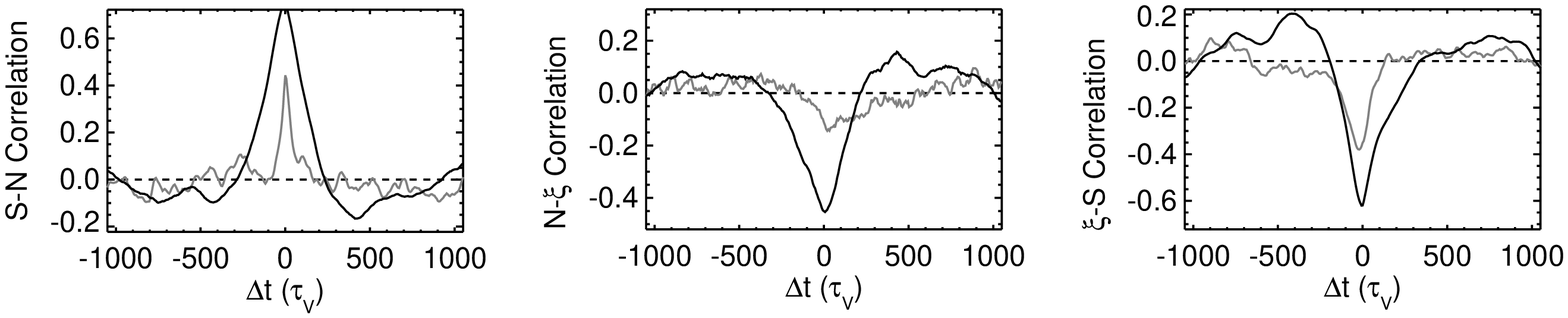, width=\linewidth}}
\caption{Cross-correlations $C(S,N), C(N,\xi), C(\xi,S)$  of spectral entropy $S(t)$, number of defects $N(t)$, and correlation length $\xi(t)$ from left to right. }
\label{f_ordercorr}
\end{figure*}

For $\epsilon = 0.08$ the quantities $S(t)$, $N(t)$, and $\xi(t)$ show smaller fluctuations than at $\epsilon = 0.17$. In fact, at higher $\epsilon$ the system visits alternately  OU and UC states, and we observed a number of instances where the system remained in the OU state for more than 500 $\tau_v$. Because ordered patterns have few defects and chaotic patterns have many, one expects these measures to be correlated with each other. Indeed, the number of defects $N$ exhibits behavior mirroring that of $S(t)$ while $\xi(t)$ is negatively correlated. Figure~\ref{f_ordercorr} shows temporal cross-correlations $C_{X,Y}(\Delta t) = \langle (X(t+\Delta t) - \langle  X\rangle) ( Y(t)  - \langle Y \rangle ) \rangle / {(\Sigma_X \Sigma_Y)}$  for $X,Y \in S(t), N(t)$, and $\xi(t)$. In all cases, the data at $\epsilon = 0.17$ is more strongly correlated.

\begin{figure*}
\centerline{\epsfig{file=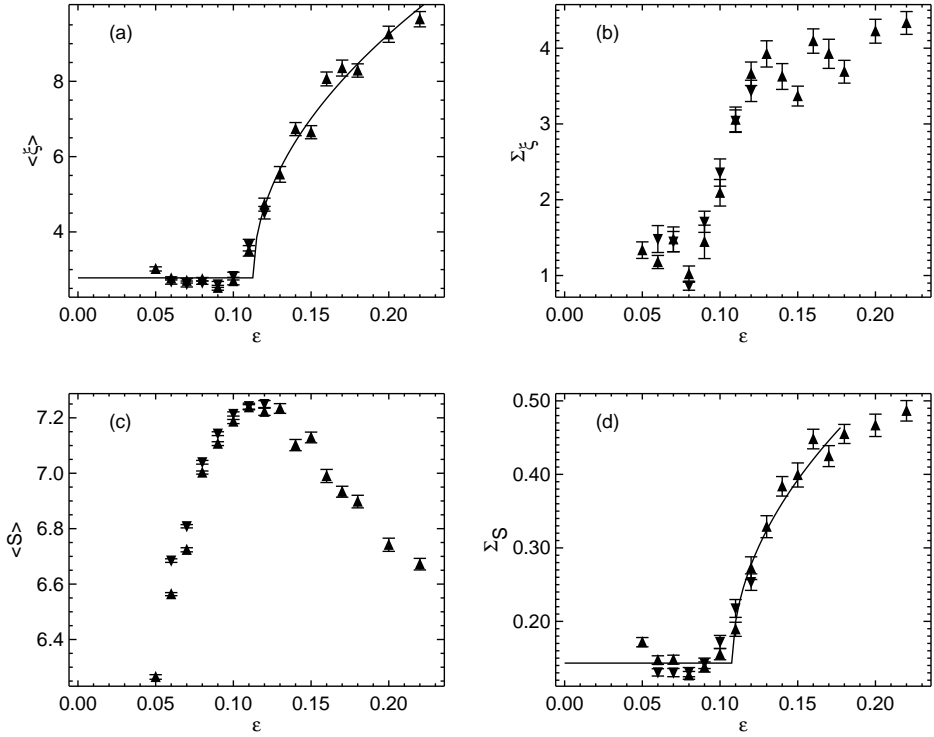, width=5in}}
\caption{(a) Mean correlation length $\langle \xi \rangle$ and (b) its standard deviation  $\Sigma_\xi$  as a function of $\epsilon$.  (c) Mean spectral entropy $\langle S \rangle$, and (d) its standard deviation $\Sigma_S$ as a function of $\epsilon$. Solid lines $\propto  \sqrt{\epsilon - \epsilon_u}$ are plotted as guides to the eye. Triangles are for short runs in which temperature steps between them were either increasing ($\blacktriangle$) or decreasing ($\blacktriangledown$).}
\label{f_ordereps}
\end{figure*}

The transition to increasing order, i.e. the appearance of long-living OU states for $\epsilon \approx 0.1 $ is a peculiar one, in that the system does not remain in the new, ordered state but intermittently returns to the chaotic state.  To characterize this transition we measured the quantities $\langle S \rangle$, $\langle \xi \rangle$, $\Sigma_S$ and $\Sigma_\xi$ at each $\epsilon$, shown in  Fig.~\ref{f_ordereps}. All four plots exhibit sudden changes near a transition point $\epsilon_t \approx 0.1$.  For instance $\langle \xi \rangle (\epsilon)$ increases sharply there, indicating that the size of ordered regions is strongly growing for $\epsilon > \epsilon_t$. Both $\Sigma_S(\epsilon)$ and $\Sigma_\xi(\epsilon)$ show that the fluctuations increase in magnitude as well when $\epsilon > \epsilon_t$. No hysteresis was observed across this transition. Note that this transition also corresponds to qualitative changes in the defect dynamics as previously described in \citet{Daniels-2002-DTI}.

\subsection{Local amplitude analysis}
\label{sec:locamp}
The analyses presented above are not specific to undulation chaos. Thus, we have used an additional method, to characterize slightly disordered undulation patterns. The idea is to use  the ansatz in Eq.~\ref{e_perfect} locally, {\em i.e.}  with  real space- and time-dependent  quantities $(q, p, A, B, C)$. The first step is to extract from  our experimental images the complex function $\psi(x,y)$  as explained at the beginning of the present  section (see Fig..~\ref{f_pics}d). Then $q(x,y)$ is obtained by the local wavenumber method described in \citet{Egolf-1998-ILP}, by calculating
\begin{equation}
q(x,y) = \sqrt{ \frac{- \partial_x^2 \psi(x,y)}{\psi(x,y)}}.
\end{equation}

Near zeros of $\psi(x,y)$ , for instance at defects, the wavevector field $q(x,y)$ field is ill-defined and yields unphysically large values, which can safely be identified and neglected. To determine the local undulation wavenumber $p(x,y)$ we use the squared modulus of $\psi$,
which can be locally well-parametrized by the following analytical expression (see Eq.~\ref{e_perfect}):
\begin{equation}
|\psi|^2 = A^2 + B^2 + C^2 - 2 B C \cos{(2py)} + 2 A (B-C) \cos{(py)}.
\label{e_bcmodulus}
\end{equation}
Consequently the local wavenumber analysis now applied to $|\psi|^2$ yields 
$p(x,y)$. For weakly disordered OU the periodicity of $|\psi|^2$ is in fact governed by $2 p$. For $B \approx C$, Eq.~\ref{e_bcmodulus} reduces to
\begin{equation}
|\psi|^2= A^2 + 2 C^2 - 2 C^2 \cos{(2py)}.
\label{e_modulus}
\end{equation}

\begin{figure*}
\centerline{\epsfig{file=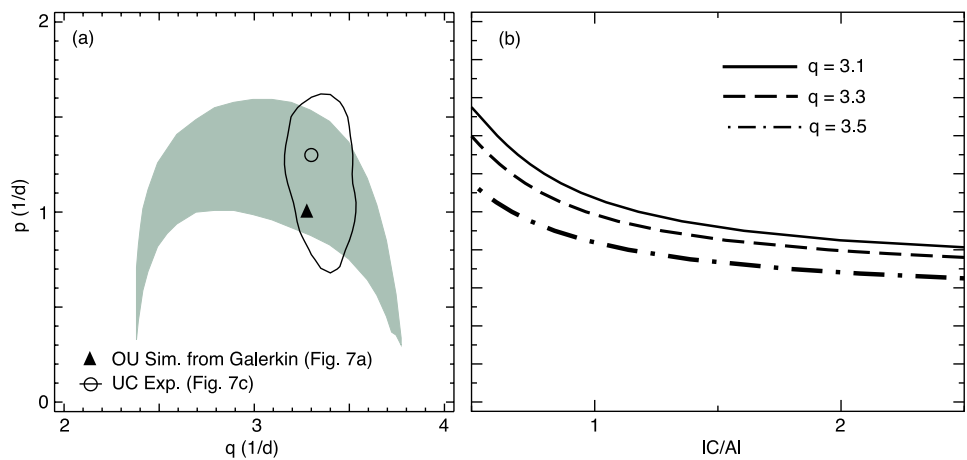, width=5in}}
\caption{(a) Stability islands at $\epsilon = 0.1$ from Fig.~\protect\ref{f_moons}, showing sample wavenumber distributions for each of the images in Fig.~\protect\ref{f_simexppics}. Peaks of the distributions are given by symbols; lines are the half-maximum of the distributions showing UC. (b) Corresponding theoretical amplitude ratios $\vert \frac{C}{A} \vert$ for $q =  3.1, 3.3, 3.5$  within the stable region. }
\label{f_balloon} 
\end{figure*}

Fig.~\ref{f_balloon} shows sample wavenumber distributions of $(q,p)$ for the experimental convection patterns cells together with the stability islands from Fig.~\ref{f_moons}. The mean values of $q, p$ (the small circle) are in the stable regime. We previously observed the mean value of $p$ to be proportional to $\sqrt{\epsilon - \epsilon_u}$, where $\epsilon_u$ agrees with the Galerkin prediction for the onset of OU \citep{Daniels-2002-DTI}. The mean roll wavenumber $q$ does not vary significantly with $\epsilon$. Inspection of Fig.~\ref{f_balloon}a shows that the distribution of the $q,p$ values overlaps with the stability regions in a wide range, and the most probable value (circle) is within the stable regime. Near defects, the undulation patterns are strongly deformed and local wavenumbers  can be  pushed outside the stability regions.

\begin{figure}
\centerline{\epsfig{file=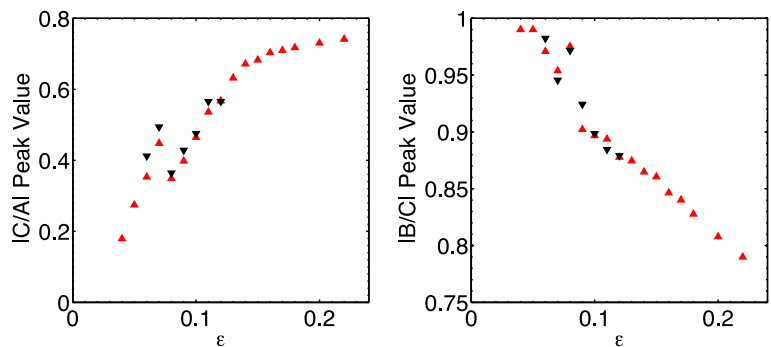, width=4in}}
\caption{Ratios $\vert \frac{C}{A} \vert$ and $\vert \frac{B}{C} \vert$ as a function of $\epsilon$, measured at the peak value of the histogram. Symbols represent averages from 500 experimental images, for which quasistatic temperature steps are increasing for $\blacktriangle$ and decreasing for $\blacktriangledown$. }
\label{f_ratioeps}
\end{figure}

\begin{figure}
\centerline{\epsfig{file=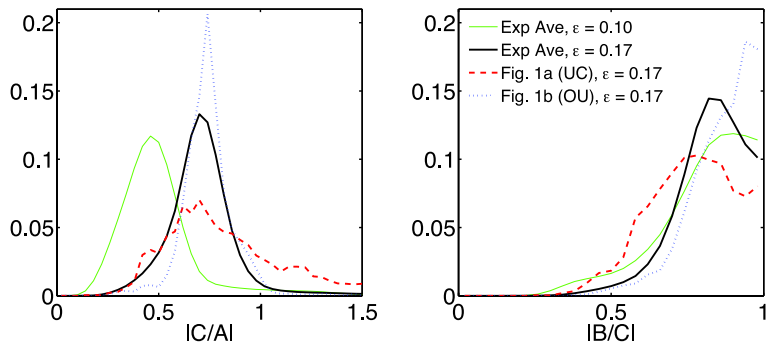, width=4in}}
\caption{Histograms of $\vert \frac{C}{A} \vert$ and  $\vert \frac{B}{C} \vert$ at two values of $\epsilon$.  Averaged experimental points (solid) are taken over 500 images, with comparison to single images (dashed, dotted) as specified in legend for $\epsilon=0.17$.}
\label{f_ratiohist}
\end{figure}

Further insight arises from distributions of the amplitudes  $A$, $B$, and $C$. For an ideal undulation pattern characterized by the wavenumbers $q, p$ we also know the amplitudes $A$ and $B=C$ according Eq.~\ref{eq:ampli}. As demonstrated in Fig.~\ref{f_balloon}b the ratio $C/A$ varies significantly with wavenumber. Thus, it is not surprising that a disordered pattern with a fairly wide range of $q, p$ values (as shown in Fig.~\ref{f_balloon}) is characterized by a broad distribution of the amplitudes $A, B \ne C$ as well.

We determined $A$, $B$, and $C$ from the local maxima and minima of the experimental $|\psi|^2$. Due to the presence of defects it is not feasible to keep track of the phases of the cosine functions in Eq.~\ref{e_bcmodulus}. Since $ y \rightarrow y = y + \pi/p$  results in $B \leftrightarrow C$ we have chosen the convention  $B < C$ in all cases. Our analysis is independent of $q(x,y)$ and $p(x,y)$ and is not sensitive to the tilt in the undulations. As shown in Fig.~\ref{f_ratioeps}, the median value of $\vert \frac{C}{A} \vert$ is observed to be low (0.2) near onset and rise with $\epsilon$. For small
$\epsilon$  the peak in $\vert \frac{B}{C} \vert$ is near the ideal OU value $B/C =1 $. With increasing $\epsilon$ the intermittently-visited UC states are responsible for an increasing asymmetry $B \ne C$, which is then reflected in the average ratio in Fig.~\ref{f_ratioeps} as well.

The averages of the ratios $C/A$ and $B/C$ shown in Fig.~\ref{f_ratioeps} are obtained from a sequence of the corresponding distribution functions examples of which are shown in Fig.~\ref{f_ratiohist}. We observe that the distributions become slightly more narrow at the higher $\epsilon$, which might indicate the increased prevalence  of OU.

One feature not relevant for the present analysis  is a weak drift of the experimental patterns down the cell \citep{Daniels-2002-DTI}. This is also the reason that undulations are mostly invisible in the time-averaged image shown in Fig.~\ref{f_homog}b. We associate the drifting behavior with non-Boussinesq effects, which combined with the effect of the basic shear render the growthrates of $\sigma_B, \sigma_C$ of the oblique modes in Eqs~.\ref{e_amplB} and \ref{e_amplC} complex.  Between $\epsilon = 0.08$ and $\epsilon = 0.17$ a drift speed of  about $0.05 d/\tau_v$ has been measured, which is is substantially smaller than the value $d/\tau_v \approx 1$ observed above $\gamma_{c2}$. The difference may be partially accounted for by the strength of the shear flow, which is greater for steeper inclinations.

\section{Discussion}

The central topic in this paper is the competition between ordered (OU) undulations and undulation chaos (UC). The Galerkin analysis of the OBE demonstrates the existence of linearly stable undulation patterns above the wavy instability up to fairly large $\epsilon$. As shown in Fig.~\ref{f_moons} the wavenumbers $q, p$, which characterize the three dominant modes $A, B, C$ (Eq.~\ref{e_perfect}) cover a finite region in $\bm q$-space, which is maximal near $\epsilon = 0.1$.
The linear stability of  OU  is confirmed by direct simulations of the OBE when starting from a Galerkin solution with noise (see Fig.~\ref{f_simexppics}a). However, the basin of attraction of the OU shrinks with increasing $\epsilon$. At small $\epsilon$ starting from random initial conditions the system selects OU (see Fig.~\ref{f_simexppics}a), albeit after a long transient. In contrast, at larger $\epsilon$ the system seems to remain in a UC state that competes with regions of OU (see Fig.~\ref{f_simexppics}b). Within the spatial  and temporal limitations of our simulation we are unable to determine whether we observe UC as a long-lived transient or as a truly asymptotically selected state. 

In the experiment we do not observe ideal OU in any regime. At small $\epsilon$, for instance at $\epsilon = 0.08$, the number of defects $N$ in our sample area (the box in Fig.~\ref{f_homog}) varies between $3$ and $18$; the average is about $N = 10$ with fluctuations between these extrema (see Fig.~\ref{f_order}b). For larger $\epsilon$ the situation is different (see Fig.~\ref{f_order}d). We find an intermittent switching between states with very few defects ($\sim4$ in Fig.~\ref{f_orderchaos}b) and others with
many (up to $N \gtrsim  25$) as in in Fig.~\ref{f_orderchaos}a. On the other hand, the system can remain for long times in either of these states, switching intermittently between them.  

We interpret this scenario as reflecting the competition between OU and UC states.
Defects distort the patterns and can easily lead to the regions of the cells where part of the the wavenumber distribution lies outside the stability regions. The creation of defects and their dynamics is a well-known mechanism allowing the system to return to the stable regime. In the optimal case, this is achieved when oppositely charged defects annihilate each other or move to the boundaries. This process takes time and is susceptible to imperfections in the cell (particularly the boundaries), but there are also cases (such as SDC, \citep{Bodenschatz-2000-RDR}) where the system never settles down to the stable attractor but switches intermittently back and forth between almost perfect ordered and strongly disordered states as in our case.


\section{Conclusion}

In this paper we have studied an anisotropic pattern forming system which shows spatiotemporal complexity. In contrast to the similar and previously studied scenario of the electrohydrodynamic instability in nematic liquid crystals \citep{Rasenat-1990-ESD}, the basic equations are simple enough to be amenable to a precise theoretical analysis. We demonstrate the crucial importance of topological defects in the competition between chaotic and well-ordered patterns. While numerical simulations demonstrate stable ordered undulations at all $\epsilon$ above the secondary instability, such a state is only intermittently accessible in experiments. Interestingly, this ordered state occurs only at {\it higher} driving, where the stability region for the undulations is largest. 

\section{Acknowledgments}

We wish to thank J.~P. Sethna and R.~J. Wiener for helpful discussions
regarding this work. We thank G. Ahlers for providing us with the code to
determine the material parameters. KED and EB acknowledge support from the National 
Science Foundation for support under grant DMR-0072077. KED thanks the 
Isaac Newton Institute for Mathematical Sciences for their hospitality and 
support which allowed the preparation of this manuscript.



\begin{thebibliography}{45}
\providecommand{\natexlab}[1]{#1}
\providecommand{\url}[1]{\texttt{#1}}
\expandafter\ifx\csname urlstyle\endcsname\relax

\bibitem[Andereck et~al.(1986)Andereck, Liu, and Swinney]{Andereck-1986-FRC}
C.~D. Andereck, S.~S. Liu, and H.~L. Swinney.
\newblock Flow regimes in a circular {C}ouette system with independently
  rotating cylinders.
\newblock \emph{Journal Of Fluid Mechanics}, 164:\penalty0 155--183, March
  1986.

\bibitem[Blondeaux(1990)]{Blondeaux-1990-SRU}
P.~Blondeaux.
\newblock Sand ripples under sea waves: 1. {R}ipple formation.
\newblock \emph{Journal Of Fluid Mechanics}, 218:\penalty0 1--17, September
  1990.

\bibitem[Bodenschatz et~al.(1988)Bodenschatz, Pesch, and
  Kramer]{Bodenschatz-1988-SDD}
E.~Bodenschatz, W.~Pesch, and L.~Kramer.
\newblock Structure and dynamics of dislocations in anisotropic pattern-forming
  systems.
\newblock \emph{Physica D}, 32\penalty0 (1):\penalty0 135--145, August 1988.

\bibitem[Bodenschatz et~al.(2000)Bodenschatz, Pesch, and
  Ahlers]{Bodenschatz-2000-RDR}
E.~Bodenschatz, W.~Pesch, and G.~Ahlers.
\newblock Recent developments in {R}ayleigh-{B}\'enard convection.
\newblock \emph{Annual Review of Fluid Mechanics}, 32:\penalty0 709--778, 2000.

\bibitem[Bottin et~al.(1998)Bottin, Daviaud, Manneville, and
  Dauchot]{Bottin-1998-DTS}
S.~Bottin, F.~Daviaud, P.~Manneville, and O.~Dauchot.
\newblock Discontinuous transition to spatiotemporal intermittency in plane
  couette flow.
\newblock \emph{Europhysics Letters}, 43:\penalty0 171--176, July 15 1998.

\bibitem[Brausch(2001)]{Brausch-thesis}
O.~Brausch.
\newblock \emph{Rayleigh-{B}\'enard {K}onvektion in verschiedenen isotropen und
  anisotropen {S}ystemen}.
\newblock PhD thesis, {U}niversit\"at {B}ayreuth, 2001.

\bibitem[Cakmur et~al.(1997)Cakmur, Egolf, Plapp, and
  Bodenschatz]{Cakmur-1997-BCS}
R.~V. Cakmur, D.~A. Egolf, B.~B. Plapp, and E.~Bodenschatz.
\newblock Bistability and competition of spatiotemporal chaotic and fixed point
  attractors in {R}ayleigh-{B}\'enard convection.
\newblock \emph{Physical Review Letters}, 79:\penalty0 1853--1856, September 8
  1997.

\bibitem[Chandrasekhar(1961)]{Chandrasekhar-1961-HHS}
S.~Chandrasekhar.
\newblock \emph{Hydrodynamic and Hydromagnetic Stability}.
\newblock Oxford University Press, Oxford, 1961.

\bibitem[Clever(1973)]{Clever-1973-FAL}
R.~M. Clever.
\newblock Finite amplitude longitudinal convection rolls in an inclined layer.
\newblock \emph{Trans. ASME C}, 95:\penalty0 407--408, August 1973.

\bibitem[Clever and Busse(1977)]{Clever-1977-ILC}
R.~M. Clever and F.~H. Busse.
\newblock Instabilities of longitudinal convection rolls in an inclined layer.
\newblock \emph{Journal of Fluid Mechanics}, 81:\penalty0 107--125, June 1977.

\bibitem[Cross and Hohenberg(1993)]{Cross-1993-PFE}
M.~C. Cross and P.~C. Hohenberg.
\newblock Pattern formation out of equilibrium.
\newblock \emph{Reviews of Modern Physics}, 65:\penalty0 851--1112, July 1993.

\bibitem[Cross and Hohenberg(1994)]{Cross-1994-SC}
M.~C. Cross and P.~C. Hohenberg.
\newblock Spatiotemporal chaos.
\newblock \emph{Science}, 263:\penalty0 1569--1570, March 1994.

\bibitem[Daniels and Bodenschatz(2002{\natexlab{a}})]{Daniels-2002-DTI}
K.~E. Daniels and E.~Bodenschatz.
\newblock Defect turbulence in inclined layer convection.
\newblock \emph{Physical Review Letters}, 88:\penalty0 034501, January 7
  2002{\natexlab{a}}.

\bibitem[Daniels and Bodenschatz(2002{\natexlab{b}})]{movie}
K.~E. Daniels and E.~Bodenschatz, 2002{\natexlab{b}}.
\newblock Movies of undulation chaos are available at through AIP EPAPS
  Document No. EPAPS: E-PRLTAO-88-001203.

\bibitem[Daniels et~al.(2000)Daniels, Plapp, and Bodenschatz]{Daniels-2000-PFI}
K.~E. Daniels, B.~B. Plapp, and E.~Bodenschatz.
\newblock Pattern formation in inclined layer convection.
\newblock \emph{Physical Review Letters}, 84:\penalty0 5320--5323, June 5 2000.

\bibitem[de~Bruyn et~al.(1996)de~Bruyn, Bodenschatz, Morris, Trainoff, Hu,
  Cannell, and Ahlers]{deBruyn-1996-ASR}
J.~R. de~Bruyn, E.~Bodenschatz, S.~W. Morris, S.~P. Trainoff, Y.~Hu, D.~S.
  Cannell, and G.~Ahlers.
\newblock Apparatus for the study of {R}ayleigh-{B}\'enard convection in gases
  under pressure.
\newblock \emph{Review of Scientific Instruments}, 67:\penalty0 2043--2067,
  June 1996.

\bibitem[Decker et~al.(1994)Decker, Pesch, and Weber]{Decker-1994-SDC}
W.~Decker, W.~Pesch, and A.~Weber.
\newblock Spiral defect chaos in {R}ayleigh-{B}\'enard convection.
\newblock \emph{Physical Review Letters.}, 73:\penalty0 648--651, August 1994.

\bibitem[Echebarria and Riecke(2000)]{Echebarria-2000-SOH}
B.~Echebarria and H.~Riecke.
\newblock Stability of oscillating hexagons in rotating convection.
\newblock \emph{Physica D}, 143:\penalty0 187--204, September 1 2000.

\bibitem[Egolf(1999)]{Egolf}
D.~A. Egolf, 1999.
\newblock private communication.

\bibitem[Egolf et~al.(1998)Egolf, Melnikov, and Bodenschatz]{Egolf-1998-ILP}
D.~A. Egolf, I.~V. Melnikov, and E.~Bodenschatz.
\newblock Importance of local pattern properties in spiral defect chaos.
\newblock \emph{Physical Review Letters}, 80:\penalty0 3228--3231, April 1998.

\bibitem[Egolf et~al.(2000)Egolf, Melnikov, Pesch, and Ecke]{Egolf-2000-MES}
D.~A. Egolf, I.~V. Melnikov, W.~Pesch, and R.~E. Ecke.
\newblock Mechanisms of extensive spatiotemporal chaos in {R}ayleigh-{B}\'enard
  convection.
\newblock \emph{Nature}, 404:\penalty0 733--736, April 13 2000.

\bibitem[Gluckman et~al.(1993)Gluckman, Marcq, Bridger, and
  Gollub]{Gluckman-1993-TAC}
B.~J. Gluckman, P.~Marcq, J.~Bridger, and J.~P. Gollub.
\newblock Time averaging of chaotic spatiotemporal wave patterns.
\newblock \emph{Physical Review Letters}, 71:\penalty0 2034--2037, September 27
  1993.

\bibitem[Gollub(1994)]{Gollub-1994-SC}
J.~P. Gollub.
\newblock Spirals and chaos.
\newblock \emph{Nature}, 367:\penalty0 318, January 1994.

\bibitem[Gollub and Cross(2000)]{Gollub-2000-NDC}
J.~P. Gollub and M.~C. Cross.
\newblock Nonlinear dynamics -- chaos in space and time.
\newblock \emph{Nature}, 404:\penalty0 710--711, April 13 2000.

\bibitem[Hansen et~al.(2001)Hansen, Hecke, Ellegaard, Andersen, Bohr, Haaning,
  and Sams]{Hansen-2001-SBT}
J.~L. Hansen, M.~Van Hecke, C.~Ellegaard, K.~H. Andersen, T.~Bohr, A.~Haaning,
  and T.~Sams.
\newblock Stability balloon for two-dimensional vortex ripple patterns.
\newblock \emph{Physical Review Letters}, 87:\penalty0 204301, November 12
  2001.

\bibitem[Hart(1971{\natexlab{a}})]{Hart-1971-SFD}
J.~E. Hart.
\newblock Stability of the flow in a differentially heated inclined box.
\newblock \emph{Journal of Fluid Mechanics}, 47:\penalty0 547--576, June
  1971{\natexlab{a}}.

\bibitem[Hart(1971{\natexlab{b}})]{Hart-1971-TWV}
J.~E. Hart.
\newblock Transition to a wavy vortex regime in convective flow between
  inclined plates.
\newblock \emph{Journal of Fluid Mechanics}, 48:\penalty0 265--271, July
  1971{\natexlab{b}}.

\bibitem[Kelly(1977)]{Kelly-1977-OND}
R.~E. Kelly.
\newblock The onset and development of {R}ayleigh-{B}\'enard convection in
  shear flows: A review.
\newblock In D.~B. Spalding, editor, \emph{Physicochemical Hydrodynamics},
  pages 65--79. Advanced Publications, 1977.

\bibitem[Kelly(1994)]{Kelly-1994-ODT}
R.~E. Kelly.
\newblock The onset and development of thermal convection in fully developed
  shear flows.
\newblock \emph{Advances in Applied Mechanics}, 31:\penalty0 35, 1994.

\bibitem[Kramer and Pesch(1995)]{Kramer-1995-CIN}
L.~Kramer and W.~Pesch.
\newblock Convection instabilities in nematic liquid-crystals.
\newblock \emph{Annual Review of Fluid Mechanics}, 27:\penalty0 515--541, 1995.

\bibitem[Kurt et~al.(2004)Kurt, Busse, and Pesch]{Kurt-2004-HCR}
E.~Kurt, F.~H. Busse, and W.~Pesch.
\newblock Hydromagnetic convection in a rotating annulus with an azimuthal
  magnetic field.
\newblock \emph{Theoretical And Computational Fluid Dynamics}, 18\penalty0
  (2-4):\penalty0 251--263, Nov 2004.

\bibitem[Morris et~al.(1993)Morris, Bodenschatz, Cannell, and
  Ahlers]{Morris-1993-SDC}
S.~W. Morris, E.~Bodenschatz, D.~S. Cannell, and G.~Ahlers.
\newblock Spiral defect chaos in large aspect ratio {R}ayleigh-{B}\'enard
  convection.
\newblock \emph{Physical Review Letters}, 71:\penalty0 2026--2029, September
  1993.

\bibitem[M\"uller et~al.(1992)M\"uller, L\"ucke, and Kamps]{Muller-1992-TCP}
H.~W. M\"uller, M.~L\"ucke, and M.~Kamps.
\newblock Transveral convection patterns in horizontal shear-flow.
\newblock \emph{Physical Review A}, 45:\penalty0 3714--3726, March 15 1992.

\bibitem[Neufeld and Friedrich(1994)]{Neufeld-1994-PFR}
M.~Neufeld and R.~Friedrich.
\newblock Pattern formation in rotating {B}\'enard convection.
\newblock \emph{International Journal of Bifurcation and Chaos in Applied
  Sciences and Engineering}, 4:\penalty0 1155--1163, October 1994.

\bibitem[Ning et~al.(1993)Ning, Hu, Ecke, and Ahlers]{Ning-1993-STA}
L.~Ning, Y.~Hu, R.~E. Ecke, and G.~Ahlers.
\newblock Spatial and temporal averages in chaotic patterns.
\newblock \emph{Physical Review Letters.}, 71:\penalty0 2216--2219, October
  1993.

\bibitem[Pesch(1996)]{Pesch-1996-CSC}
W.~Pesch.
\newblock Complex spatiotemporal convection patterns.
\newblock \emph{Chaos}, 6:\penalty0 348--357, September 1996.

\bibitem[Ramazza et~al.(1992)Ramazza, Residori, Giacomelli, and
  Arecchi]{Ramazza-1992-STD}
P.~L. Ramazza, S.~Residori, G.~Giacomelli, and F.~T. Arecchi.
\newblock Statistics of topological defects in linear and nonlinear optics.
\newblock \emph{Europhysics Letters}, 19:\penalty0 475--80, July 6 1992.

\bibitem[Rasenat et~al.(1990)Rasenat, Steinberg, and Rehberg]{Rasenat-1990-ESD}
S.~Rasenat, V.~Steinberg, and I.~Rehberg.
\newblock Experimental studies of defect dynamics and interaction in
  electrohydrodynamic convection.
\newblock \emph{Physical Review A}, 42\penalty0 (10):\penalty0 5998--6008,
  November 1990.

\bibitem[Rehberg et~al.(1989)Rehberg, Rasenat, and Steinberg]{Rehberg-1989-TWD}
I.~Rehberg, S.~Rasenat, and V.~Steinberg.
\newblock Traveling waves and defect-initiated turbulence in electroconvecting
  nematics.
\newblock \emph{Physical Review Letters}, 62:\penalty0 756--9, February 7 1989.

\bibitem[Ruth(1980)]{Ruth-1980-TTR}
D.~W. Ruth.
\newblock On the transition to transverse rolls in inclined infinite fluid
  layers-steady solutions.
\newblock \emph{International Journal of Heat and Mass Transfer}, 23:\penalty0
  733--737, May 1980.

\bibitem[Ruth et~al.(1980)Ruth, Raithby, and Hollands]{Ruth-1980-SII}
D.~W. Ruth, G.~D. Raithby, and K.~G.~T. Hollands.
\newblock On the secondary instability in inclined air layers.
\newblock \emph{Journal of Fluid Mechanics}, 96:\penalty0 481--492, February
  1980.

\bibitem[Tagg(1994)]{Tagg-1994-CTP}
R.~Tagg.
\newblock The {C}ouette-{T}aylor problem.
\newblock \emph{Nonlinear Science Today}, 4:\penalty0 1--24, 1994.

\bibitem[Trainoff and Cannell(2002)]{Trainoff-2002-POT}
S.~P. Trainoff and D.~S. Cannell.
\newblock Physical optics treatment of the shadowgraph.
\newblock \emph{Physics Of Fluids}, 14:\penalty0 1340--1363, April 2002.

\bibitem[Young and Riecke(2003)]{Young-2003-PHD}
Y.~N. Young and H.~Riecke.
\newblock Penta-hepta defect chaos in a model for rotating hexagonal
  convection.
\newblock \emph{Physical Review Letters}, 90:\penalty0 134502, 2003.

\bibitem[Yu et~al.(1997)Yu, Chang, and Lin]{Yu-1997-SMT}
C.~H. Yu, M.~Y. Chang, and T.~F. Lin.
\newblock Structures of moving transverse and mixed rolls in mixed convection
  of air in a horizontal plane channel.
\newblock \emph{International Journal of Heat and Mass Transfer}, 40\penalty0
  (2):\penalty0 333--46, January 1997.

\end{thebibliography}

\end{document}